\documentclass[prl,twocolumn,aps,showpacs]{revtex4}
\usepackage{graphicx}
\begin{document}

\title{Spin-3 Chromium Bose-Einstein Condensates}
\author{L. Santos$^{(1)}$ and T. Pfau$^{(2)}$}
\affiliation{
(1) Institut f\"ur Theoretische Physik III, 
Universit\"at Stuttgart, Pfaffenwaldring 57 V, D-70550 Stuttgart \\
(2) 5. Physikalisches Institut, 
Universit\"at Stuttgart, Pfaffenwaldring 57 V, D-70550 Stuttgart \\
}

\begin{abstract}  
We analyze the physics of spin-3 Bose-Einstein condensates, and in particular the 
new physics expected in on-going experiments with condensates 
of Chromium atoms. We first discuss the ground-state properties, which, depending 
on still unknown Chromium parameters, and for low magnetic fields 
can present various types of phases. 
We also discuss the spinor-dynamics in Chromium spinor condensates, which 
present significant qualitative differences when compared to other 
spinor condensates. In particular, dipole-induced spin relaxation may lead 
for low magnetic fields to transfer of spin into angular momentum similar to the 
well-known Einstein-de Haas effect. Additionally, a rapid large transference 
of population between distant magnetic states becomes also possible. 
\end{abstract}  
\maketitle


Within the very active field of ultra cold atomic gases, 
multicomponent (spinor) Bose-Einstein condensates (BECs) have recently 
attracted a rapidly growing attention. Numerous works have addressed 
the rich variety of phenomena revealed by spinor BEC, in particular 
in what concerns ground-state and spin dynamics  
\cite{Ho98,Ohmi98,Law98,Koashi00,Ciobanu00,Koashi00b,Pu01,You04,Ho00b,Hall98,
Stenger98,Barret01,Schmaljohann04,Chang04,Kuwamoto04,Wheeler04,Higbie05,Widera05}.
The first experiments on spinor BEC were performed at JILA using mixtures of 
$^{87}$Rb BEC in two magnetically confined internal states (spin-$1/2$ BEC) 
\cite{Hall98}. Optical trapping of spinor condensates was first realized
in spin-$1$ Sodium BEC at MIT  \cite{Stenger98} constituting a major
breakthrough since, whereas magnetic trapping confines the BEC to weak-seeking 
magnetic states, an optical trap enables confinement of all magnetic
substates. In addition, the atoms in a magnetic substate can be converted into
atoms in other substates through interatomic interactions. Hence, these
experiments paved the way towards the above mentioned fascinating 
phenomenology  
originating from the spin degree of freedom. Various experiments have been
realized since then in spin-$1$ BEC, in particular in $^{87}$Rb in the $F=1$
manifold. It has been predicted that spin-$1$ BEC can present just two
different ground states phases, either ferromagnetic or polar \cite{Ho98}.   
In the case of $F=1$ $^{87}$Rb it has been shown that the ground-state 
presents a ferromagnetic behavior \cite{Schmaljohann04,Chang04,Klausen01}. 
These analyses have 
been recently extended to spin-$2$ BEC, which presents an even richer 
variety of possible ground-states, including in addition the 
so-called cyclic phases \cite{Ciobanu00,Koashi00b}. Recent experiments have
shown a behavior compatible with a polar ground-state, although in the very
vicinity of the cyclic phase \cite{Schmaljohann04,Klausen01}. Recently, spin
dynamics has attracted a major interest, revealing the 
fascinating physics of the coherent oscillations between the different 
components of the manifold \cite{Barret01,Schmaljohann04,Chang04,Kuwamoto04,
Wheeler04,Higbie05}. 


Very recently, a  Chromium BEC (Cr-BEC) has been achieved at Stuttgart University
\cite{Griesmaier05}. Cr-BEC presents fascinating new features when
compared to other experiments in BEC. On one hand, since the ground state of 
$^{52}$Cr is $^7$S$_3$, Cr-BEC constitutes the first accessible example 
of a spin-3 BEC. We  show below that this fact may have very important
consequences for both the ground-state and the dynamics of spinor Cr-BEC. 
On the other hand, when aligned into the state with magnetic quantum number $m=\pm 3$, 
$^{52}$Cr presents a magnetic moment $\mu=6\mu_B$, where $\mu_B$ is
the Bohr magneton. This dipolar moment 
should be compared to alkali atoms, which have a
maximum magnetic moment of $1\mu_B$, and hence $36$ times smaller
dipole-dipole interactions. Ultra cold dipolar gases have attracted a 
rapidly growing attention, in particular in what 
concerns its stability and excitations \cite{Dipoles}. The interplay of the 
dipole-dipole interaction and spinor-BEC physics has been also considered 
\cite{Pu01,You04}. Recently, the dipolar
effects have been observed for the first time in 
the expansion of a Cr-BEC \cite{Stuhler05}.


This Letter analyzes spin-3 BEC, and in particular 
the new physics expected in on-going experiments in Cr-BEC. After deriving the
equations that describe this system, we focus on the ground-state, using single-mode approximation (SMA), showing 
that various phases are possible, depending on the applied magnetic field, 
and the (still unknown) value of the $s$-wave scattering length for the channel of total spin zero. 
This phase diagram presents certain differences with respect to the diagram first worked out recently by Diener and Ho \cite{Diener05}.  
In the second part of this Letter, we discuss the spinor dynamics, departing from the SMA. 
The double nature of Cr-BEC as a spin-3 BEC and a dipolar BEC is shown to lead to significant qualitative
differences when compared to other spinor BECs. The larger spin 
can allow for fast population transfer from $m=0$ to $m=\pm 3$ without a
sequential dynamics as in $F=2$ $^{87}$Rb \cite{Schmaljohann04}. In addition, dipolar
relaxation violates spin conservation, leading to rotation of the
different components, resembling the well-known Einstein-de Haas (EH) 
effect~\cite{Einstein15}.


In the following we  consider an optically trapped Cr-BEC with 
$N$ particles. The Hamiltonian regulating Cr-BEC is of the form 
$\hat H=\hat H_0+\hat V_{sr}+\hat V_{dd}$. 
The single-particle part of the Hamiltonian, $\hat H_0$, includes the 
trapping energy and the linear Zeeman effect (quadratic Zeeman 
effect is absent in Cr-BEC), being of the form
\begin{equation}
\hat H_0=\int d{\bf r} \sum_m \hat\psi_m^\dag({\bf r})
\left [\frac{-\hbar^2}{2M}\nabla^2+U_{trap}({\bf r})+pm \right ]
\hat\psi_m({\bf r}),
\end{equation}
where $\hat \psi_m^\dag$ ($\hat\psi_m$) is the creation (annihilation) 
operator in the $m$ state, $M$ is the atomic mass, and $p=g\mu_B B$, with $g=2$ for $^{52}$Cr, 
and $B$ is the applied magnetic field.

The short-range (van der Waals) interactions are given by $\hat V_{sr}$. 
For any interacting pair with spins ${\bf S}_{1,2}$, $\hat V_{sr}$ 
conserves the total spin, $S$, and 
is thus described in terms of the projector operators on different 
total spins $\hat{\cal P}_S$, where $S=0$, $2$, $4$, and $6$ 
(only even $S$ is allowed)\cite{Ho98} :
\begin{equation}
\hat V_{sr}=\frac{1}{2}\int d{\bf r} \sum_{S=0}^6 g_S \hat{\cal P}_S({\bf r}),
\end{equation}
where $g_S=4\pi\hbar^2a_S/M$, and $a_S$ is the $s$-wave scattering length for a 
total spin $S$. Since ${\bf S}_1\cdot{\bf S}_2=(S^2-S_1^2-S_2^2)/2$, then
$\sum_S \hat{\cal P}_S({\bf r}) = :\hat n^2({\bf r}):$, 
$\sum_S \lambda_S \hat{\cal P}_S({\bf r}) = :\hat F^2({\bf r}):$, 
and 
$\sum_S \lambda_S^2 \hat{\cal P}_S({\bf r}) = :\hat O^2({\bf r}):$, 
where $::$ denotes normal order, 
$\lambda_S=[S(S+1)-24]/2$, 
$\hat n({\bf r})=\sum_m \hat\psi_m^\dag({\bf r})\hat\psi_m({\bf r})$, 
$\hat F^2=\sum_{i=x,y,z} \hat F_i^2$, with 
$\hat F_{i}({\bf r})=\sum_{mn} \hat\psi_m^\dag({\bf r})S_{mn}^i\hat\psi_n({\bf r})$, 
and $\hat O^2=\sum_{i,j} \hat O_{ij}^2$, with 
$\hat O_{ij}=\sum_{m,n} \hat\psi_m^\dag({\bf r})(S^iS^j)_{mn}\hat\psi_n(\vec
r)$. $S^{x,y,z}$ are the spin matrices.
Hence,
\begin{equation}
\hat V_{sr}=\frac{1}{2}\int d{\bf r} [c_0 :\hat n^2({\bf r}) + 
c_1 \hat F^2({\bf r}) +c_2 \hat{\cal P}_0({\bf r})+c_3 \hat O^2({\bf r}):],
\label{Vsr}
\end{equation}
where 
$\hat {\cal P}_0({\bf r})=\frac{1}{7}\sum_{m,n}(-1)^{m+n}
\hat\psi_m^\dag\hat\psi_{-m}^\dag\hat\psi_n\hat\psi_{-n}$, 
and
$c_0=(-11g_2+81g_4+7g_6)/77$, 
$c_1=(g_6-g_2)/18$, 
$c_2=g_0+(-55g_2+27g_4-5g_6)/33$,
$c_3=g_2/126-g_4/77+g_6/198$.
For the case of $^{52}$Cr \cite{Werner05}, $a_6=112a_B$, where $a_B$ 
is the Bohr radius, and 
$c_0=0.65g_6$, $c_1=0.059g_6$, $c_2=g0+0.374g_6$, and $c_3=-0.002g_6$.
The value of $a_0$ is unknown, and 
hence, in the following, we  consider different scenarios depending on the
value of $g_0/g_6$. Note that Eq.~(\ref{Vsr}) is similar to that obtained 
for spin-2 BEC \cite{Ciobanu00,Koashi00b},  
the main new feature being the $c_3$ term, which 
introduce qualitatively new physics as discussed below.

The dipole-dipole interaction $\hat V_{dd}$ is of the form
\begin{eqnarray}
\hat V_{dd}&=&\frac{c_d}{2}\int d{\bf r}\int d{\bf r}' \frac{1}{|{\bf r}-{\bf r}'|^3}
\hat\psi_m^\dag ({\bf r})\hat\psi_m'^\dag ({\bf r}') \nonumber \\
&&\!\!\!\!\!\!\!\!\!\!\!\!\!\!\!\! \!\!\!\!\left [
{\bf S}_{mn}\cdot{\bf S}_{m'n'}-3({\bf S}_{mn}\cdot{\bf e})
({\bf S}_{m'n'}\cdot{\bf e})
\right ]
\hat\psi_n ({\bf r})\hat\psi_n'({\bf r}'),
\end{eqnarray}
where $c_d=\mu_0\mu_B^2g_S^2/4\pi$, with $\mu_0$ the magnetic permeability of 
vacuum, and ${\bf e}=({\bf r}-{\bf r}')/|{\bf r}-{\bf r}'|$. For $^{52}$Cr $c_d=0.004g_6$. 
$\vec V_{dd}$ may be re-written as:
\begin{eqnarray}
\hat V_{dd}&=&\sqrt{\frac{3\pi}{10}}c_d\int \int \frac{d{\bf r} d{\bf r}'}{|{\bf r}-{\bf r}'|^3} 
:\left \{ 
{\cal F}_{zz}({\bf r},{\bf r}') Y_{20}
\right\delimiter 0 \nonumber \\
&&\left\delimiter 0 +{\cal F}_{z,-}({\bf r},{\bf r}') Y_{21}
+ {\cal F}_{z,+}({\bf r},{\bf r}') Y_{2-1} \right\delimiter 0
\nonumber \\
&&\left\delimiter 0 
+ {\cal F}_{-,-}({\bf r},{\bf r}')Y_{22}
+ {\cal F}_{+,+} ({\bf r},{\bf r}') Y_{2-2}
\right\}:,
\end{eqnarray}
where ${\cal F}_{zz}({\bf r},{\bf r}')=\sqrt{2/3} [3\hat F_z({\bf r})\hat F_z({\bf r}')-
\hat {\bf F}({\bf r})\cdot \hat {\bf F}({\bf r}')]$, 
${\cal F}_{z,\pm}({\bf r},{\bf r}')=\pm[\hat F_\pm({\bf r})\hat F_z({\bf r}')+
\hat F_z({\bf r})\hat F_\pm({\bf r}')]$, 
${\cal F}_{\pm,\pm}({\bf r},{\bf r}')=\hat F_\pm({\bf r})\hat F_\pm({\bf r}')$, 
$\hat F_\pm=\hat F_x\pm i\hat F_y$, and $Y_{2m}({\bf r}-{\bf r}')$ are the spherical
harmonics. Note that contrary to the short-range interactions, $\hat V_{dd}$ does 
not conserve the total spin, and 
may induce a transference of angular momentum into the center of mass (CM) degrees of freedom.

We  first discuss the ground-state of the spin-3 BEC for 
different values of $g_0$, and the 
magnetic field, $p$. We  consider mean-field (MF) 
approximation $\hat\psi_m({\bf r})\simeq\sqrt{N}\psi_m({\bf r})$. 
In order to simplify the analysis of the possible 
ground-state solutions we  perform 
SMA: $\psi_m({\bf r})=\Phi({\bf r})\psi_m$, 
with $\int d{\bf r} |\Phi({\bf r})|^2=1$. 
Apart from spin-independent terms the energy per particle is of the form:
\begin{eqnarray}
E&=& pf_z+\frac{Nc_1\beta}{2}(f_z^2+f_+f_-) +\frac{2c_2N\beta}{7}|s_-|^2
\nonumber \\
&+& \frac{Nc_3\beta}{2}\sum_{ij}O_{ij}^2 +
\sqrt{\frac{3\pi}{10}}Nc_d 
\left [
\sqrt{2/3}
\Gamma_0 (2f_z^2-f_+f_-) \right\delimiter 0 \nonumber \\
&-&\left\delimiter 0 2\Gamma_{+1}f_zf_-+2\Gamma_{-1}f_zf_+ 
+\Gamma_{+2}f_-^2+\Gamma_{-2}f_+^2
\right ]
,
\end{eqnarray}
where $\beta=\int d{\bf r} |\Phi({\bf r})|^4$, 
$s_-=\frac{1}{2}\sum_m (-1)^m \psi_m\psi_{-m}$, 
$f_i=\sum_{m,n} \psi_m^* S^i_{mn} \psi_n$,
$O_{ij}=\sum_{m,n} \psi_m^* (S^iS^j)_{mn} \psi_n$
and $\Gamma_{m}=\int d{\bf r}\int d{\bf r}' 
|\Phi({\bf r}')|^2|\Phi({\bf r})|^2 Y_{2m}({\bf r}-{\bf r}')/|{\bf r}-{\bf r}'|^3$. 
Note that $\Gamma_{\pm 1}=0$ for any symmetric density $|\Phi({\bf r})|^2$.
\begin{figure}
\begin{center}\
\vspace*{-0.5cm}
\includegraphics[width=6.2cm,angle=0]{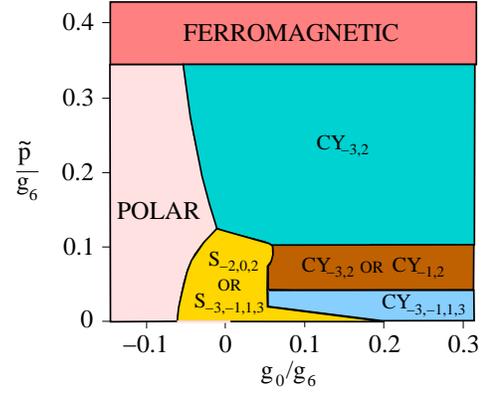}
\end{center} 
\caption{Phase diagram function of $g_0/g_6$ and 
$\tilde p/g_6$, where $\tilde p=2g_S\mu_BB/N\beta$.}
\label{fig:1}  
\end{figure}
 
Let us consider a magnetic field in the $z$-direction. Hence, 
the ground-state magnetization must be aligned with the external field, and 
$f_+=f_-=0$. Then, we obtain that 
$\sum_{ij}O_{ij}^2=77-12\gamma_z+3\gamma_z^2/2-f_z^2/2+|\eta|^2/2+2|\sigma|^2$, 
where $\gamma_z =\sum_m m^2 |\psi_m|^2$, 
$\eta=\sum_{m=-3}^1 \sqrt{[12-(m+2)(m+1)][12-m(m+1)]} \psi_{m+2}^*\psi_m$, and 
$\sigma=\sum_{m=-3}^2  m\sqrt{[12-m(m+1)]} \psi_{m+1}^*\psi_m$. 
Removing spin-independent terms, the energy becomes $E=N\beta\epsilon/2$, with 
\begin{equation}
\epsilon=\tilde p f_z +\tilde c_1 f_z^2+\frac{4c_2}{7}|s_-|^2+
c_3 \left ( \frac{3\gamma_z^2}{2}-12\gamma_z +\frac{|\eta|^2}{2}+2|\sigma|^2 \right),
\label{eps}
\end{equation}
where $\tilde p=2p/N\beta$, and $\tilde c_1=c_1-c_3/2+\sqrt{16\pi/5}\Gamma_0
c_d/\beta$. Since $c_d\ll c_1$, dipolar effects are not relevant 
for the equilibrium discussion. We will hence set $\Gamma_0=0$.


We have minimized Eq.~(\ref{eps}) with respect 
to $\psi_m$, under the constraints $\sum_m|\psi_m|^2=1$ and $f_+=0$ \cite{footnote1}. 
Figs.~\ref{fig:1} shows the corresponding phase diagram, which although in basic agreement with that worked out recently in Ref.~\cite{Diener05} 
presents certain differences in its final form. For the phases discussed below $\sigma=0$ \cite{footnoteZ}. For sufficiently negative $g_0$ the system is in a polar phase
$P=({\bf c}\theta,0,0,0,0,0,{\bf s}\theta)$, with ${\bf c}\equiv\cos$, and ${\bf s}\equiv\sin$. This phase 
is characterized by $f_z\simeq -3\tilde p/\tilde p_c$ ($p_c\simeq 6\tilde c_1\simeq 0.36$), $4|s_-|^2=1-(\tilde p/\tilde p_c)^2$, 
$\gamma_z=9$, and $|\eta|=0$. The polar phase extends up to $g_0/g_6=0.01$ for $p=0$.
For sufficiently large $g_0$ and $\tilde p$, $CY_{-3,2}=({\bf c}\theta,0,0,0,0,{\bf s}\theta,0)$ 
occurs. This phase is cyclic ($s_-=0$) and it is characterized 
by $f_z=-3\tilde p/\tilde p_c$, $\gamma_z\simeq 6+3\tilde p/\tilde p_c$, and $|\eta|=0$. Both P and $CY_{-3,2}$ continuously 
transform at $\tilde p=\tilde p_c$ into a ferromagnetic phase $F=(1,0,0,0,0,0,0)$. For $0.04<\tilde p\leq 2\tilde c_1+9c_3$ 
$CY_{-3,2}$ becomes degenerated with the cyclic phase 
$CY_{-1,2}=(0,0,{\bf c}\theta,0,0,0,{\bf s}\theta,0)$. The latter state differs 
from $CY_{-3,2}$, since $\gamma_z=2-3\tilde p/\tilde p_c$. For sufficiently large $g_0$ and $\tilde p<0.04$, 
a cyclic phase ($CY_{-3,-1,1,3}$) of the form $(\psi_{-3},0,\psi_{-1},0,\psi_{1},0,\psi_{-3})$ occurs. Contrary to the 
other cyclic phases this phase is characterized by $|\eta|>0$ \cite{FootnoteBiaxial}. Finally for a region around $g_0=0$ 
an additional phase with $|\eta|>0$, $|s_-|>0$ is found. 
In this phase two possible ground states are degenerated \cite{footnoteG}, 
namely $S_{-2,0,2}=(0,\psi_{-2},0,\psi_{0},0,\psi_{+2},0)$, and $S_{-3,-1,1,3}$, which has a similar form as $CY_{-3,-1,1,3}$.

For a Cr-BEC in a spherical trap of frequency $\omega$ 
in the Thomas-Fermi regime, $\tilde p=\tilde p_c$ for a magnetic field (in mG) 
$B\simeq 1.25
\left (
N/10^5
\right )^{2/5}
\left (
\omega/10^3\times 2\pi
\right )^{6/5}
$. Hence, most probably, magnetic shielding seems necessary for the observation of 
non-ferromagnetic ground-state phases. We would like to stress as well, that there are 
still uncertainties in the exact values of $a_{2,4,6}$. In this sense, we have checked for $a_6=98a_B$, $a_4=64a_B$, and 
$a_2=-27a_B$, that apart from a shift of $0.2$ towards larger values of $g_0/g_6$, the phase diagram remains 
qualitatively the same.

In the last part of this Letter, we  consider the spinor dynamics 
within the MF approximation, but abandoning the 
SMA. The equations for the dynamics are obtained by deriving 
the MF Hamiltonian 
with respect to $\psi_m^*({\bf r})$:
\begin{eqnarray}
i\hbar\frac{\partial}{\partial t}\psi_m({\bf r})&=&
\left [  
\frac{-\hbar^2\nabla^2}{2M}+U_{trap}+pm\right] \psi_m
\nonumber \\
&+& N
\left [
c_0 n+ m(c_1 f_z+c_d {\cal A}_0) \right ] \psi_m \nonumber \\
&+&\frac{N}{2}
\left[c_1 f_-+ 2c_d {\cal A}_{-} \right ] S^+_{m,m-1} \psi_{m-1}\nonumber \\
&+& \frac{N}{2} 
\left[c_1f_+ + 2c_d {\cal A}_{+}\right ] S^-_{m,m+1}\psi_{m+1}\nonumber \\
&+& (-1)^m \frac{2N c_2}{7}s_-\psi_{-m}^* \nonumber \\
&+& N c_3 \sum_{n} \sum_{i,j}O_{ij}(S^i S^j)_{mn}\psi_n,
\label{dynamic}
\end{eqnarray}
where 
${\cal A}_0=\sqrt{6\pi/5} [\sqrt{8/3}\Gamma_{0,z}+\Gamma_{1,-}+\Gamma_{-1,+}]$,
${\cal A}_\pm=\sqrt{6\pi/5} [-\Gamma_{0,\pm}/\sqrt{6}\mp\Gamma_{\pm,z}+\Gamma_{\pm 2,\mp}]$, 
$\Gamma_{m,i}=\int d{\bf r}'  f_i({\bf r}') 
Y_{2m}({\bf r}-{\bf r}')/|{\bf r}-{\bf r}'|^3$, and 
$S^\pm_{m,m\mp 1}=\sqrt{12-m(m\mp 1)}$. 
Note that all $\psi_m$, $n$, $f_i$, $O_{ij}$, and ${\cal A}_i$ have now a spatial dependence. 
We first consider $p=0$, and discuss on $p\neq 0$ below.
There are two main features in the spinor dynamics in Cr-BEC  
which are absent (or negligible) in other spinor BECs. 
On one side the $c_3$ term allows for jumps in the spin manifold 
larger than one, and hence for a rapid dynamics from e.g. $m=0$ 
to $m=\pm 3$. On the other side, 
the dipolar terms induce a EH-like transfer of spin into CM 
angular momentum.
\begin{figure}[ht] 
\begin{center}\
\vspace*{-0.6cm}
\hspace*{-1.2cm}
\includegraphics[width=10.5cm,angle=0]{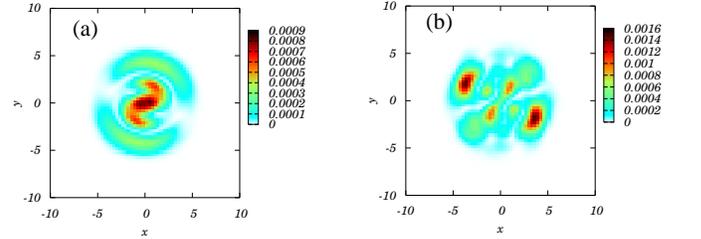}
\end{center} 
\caption{$|\psi_{m=-2}({\bf r})|^2$  at $\omega t=40$ (a) and $120$ (b) for  
$p=0$, $g_0=0$, $\omega_z=1$kHz, $N=10^4$ atoms, and  $\psi(t=0)=\psi_{m=-3}$.  
The $x$ and $y$ axes are in $\sqrt{\hbar/m\omega}$ units.}
\label{fig:2}  
\end{figure}

We consider for simplicity a quasi-2D BEC, i.e. 
a strong confinement in the $z$-direction by a harmonic potential 
of frequency $\omega_z$. Hence 
$\psi_m({\bf r})=\phi_0(z)\psi_m({\bf \rho})$, where $
\phi_0(z)=\exp[-z^2/2l_z^2]/\pi^{1/4}\sqrt{l_z}$,
with $l_z=\sqrt{\hbar/m/\omega_z}$. We have then 
solved the 2D equations using Crank-Nicholson method, considering a harmonic 
confinement of frequency $\omega$ in the $xy$-plane. 
In 2D, $\Gamma_{\pm 1,z}=0$, but these terms 
vanish also in 3D due to symmetry, and hence the 2D physics 
is representative of the 3D one. The vanishing of $\Gamma_{\pm 1,z}$ is rather important, 
since if
$\psi(t=0)=\psi_{m=\pm 3}$, $\Gamma_{\pm 1,z}$ 
is responsible for a fast dipolar relaxation (for $\omega t\sim 1$). 
Hence, a BEC with $\psi(t=0)=\psi_{\pm 3}$ does not 
present any significant fast spin dynamics. 
However significant spin relaxation appears in the
long time scale, due to the $\Gamma_{2,-}({\bf r})$ terms.  
This is the case of Fig.~\ref{fig:2}, where we $\psi(t=0)=\psi_{m=-3}$ \cite{footnote2}.

One of the most striking effects related with spin relaxation 
is the transference of spin into CM angular 
momentum, which resembles the famous EH effect \cite{Einstein15}. 
The analysis of this effect motivated us to avoid the SMA in the 
study of the spinor dynamics. In Figs.~\ref{fig:2} we 
show snapshots of the spatial distribution of 
the $m=-2$ component. 
Observe that the wavefunction clearly loose its polar symmetry, since spin is converted 
into orbital angular momentum. The spatial patterns become progressively more complicated in time.

\begin{figure}
\begin{center}\
\vspace*{-0.6cm}
\includegraphics[width=5.8cm,angle=0]{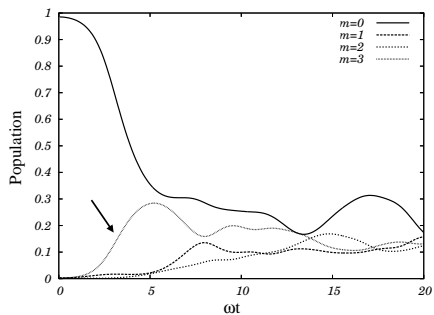}
\end{center} 
\caption{Population of $\psi_{0,1,2,3}$ versus $\omega t$ for $g_0=g_6$, $p=0$, and 
$\psi (0)=\psi_{0}$. Note the rapid growth of $m=\pm 3$ (arrow).} 
\label{fig:3}  
\end{figure}

The other special feature of Cr-BEC, namely the appearance of the $c_3$ term  
can have significant qualitative effects in the dynamics both for short and
for long time scales. The evolution at long time scale may present interesting
features, as large revivals, and it will be considered in future work. Here,
we would like to focus on the short-time scales, where the $c_3$ term 
may produce fast transference from $\psi_{m=0}$ 
to the extremes $\psi_{m=\pm 3}$. The latter is illustrated in Fig.~\ref{fig:3}, where we consider $g_0=g_6$. 
The population is initially
all in the $m=0$ \cite{footnote2}. Contrary to the case of $F=2$ in $^{87}$Rb 
\cite{Schmaljohann04}, there is at short time scales a jump to the extremes (the 
population of $\pm m$ is the same due to symmetry since $p=0$).
This large jump is absent if $c_3=0$, and depends on the value of $g_0$. 
In particular, if $g_0=0$ 
one obtains at short time scales a sequential population as for $F=2$ $^{87}$Rb 
\cite{Schmaljohann04}.

We finally comment on the dynamics if $p\neq 0$. In the case of 
$F=1$ or $F=2$ $^{87}$Rb, the dynamics is independent of $p$ since the 
linear Zeeman effect may be gauged out by transforming  
$\psi_m\rightarrow\psi_m e^{ipmt/\hbar}$, due to the conservation of the
total spin. In Cr-BEC the situation is very different, since 
the $\Gamma_{\pm 1,z}$ and $\Gamma_{\pm 2,\mp}$ do not conserve the total spin, 
and hence oscillate with the Larmor frequency $\omega_L=p/\hbar$ and
$2\omega_L$, respectively. If $\omega_L\gg \omega$ one may perform rotating-wave
approximation and eliminate these terms. Hence 
the coherent EH-like effect disappears for sufficiently large applied magnetic fields.

In conclusion, spin-3 Cr-BEC is predicted to show different types of spin phases 
depending on $a_0$ and the magnetic field. 
The spinor dynamics also presents novel features, as a fast transference 
between $\psi_0\rightarrow\psi_{\pm 3}$, and the Einstein-de Haas-like transformation of spin into rotation of 
the different components due to the dipole interaction.

We would like to thank M. Fattori for enlightening discussions,
and the German Science Foundation (DFG) (SPP1116 and
SFB/TR 21) for support. We thank H. M\"akel\"a and K.-A. Suominen for pointing us a mistake 
in previous calculations, and T.-L. Ho for enlightening e-mail exchanges. 
During the elaboration of the final version of this paper, 
the EH-effect has been also discussed in Ref.~\cite{Kawaguchi05}.


\begin{thebibliography}{99}
\bibitem{footnoteZ} We also found numerically states (similar to the Z states of Ref.~\cite{Diener05}) 
with tiny $\sigma$, which are however very slight variations of neighboring phases, 
with which they are in practice degenerated.
\bibitem{footnoteG} These phases are degenerated for any practical purposes, with relative energy differences  $<0.01\%$.
\end{thebibliography}
\end{document}